\documentclass[11pt]{article}

\usepackage{graphics,amsmath,amssymb,amsthm,accents}
\usepackage{tipa}
\usepackage{epic}
\usepackage{amsfonts}
\usepackage{accents}
\usepackage{texdraw}
\usepackage{color}
\usepackage{eucal}
\usepackage{cite}
\usepackage{bm}
\newtheorem{Proposition}{Proposition}
\numberwithin{equation}{section}

\def \tyb#1{\hbox{\tiny{[{\it{#1}}]}}}
\def \ty#1{\hbox{\tiny{{\it{#1}}}}}

\def \S#1{S^{(#1)}}

\DeclareMathAccent{\wtilde}{\mathord}{largesymbols}{"65}
\DeclareMathAccent{\what}{\mathord}{largesymbols}{"62}

\def\m@th{\mathsurround=0pt}
\mathchardef\bracell="0365
\def\upbrall{$\m@th\bracell$}
\def\undertilde#1{\mathop{\vtop{\ialign{##\crcr
    $\hfil\displaystyle{#1}\hfil$\crcr
     \noalign
     {\kern1.5pt\nointerlineskip}
     \upbrall\crcr\noalign{\kern1pt
   }}}}\limits}

\def\underhat#1{\mathop{\vtop{\ialign{##\crcr
    $\hfil\displaystyle{#1}\hfil$\crcr
     \noalign
     {\kern1.5pt\nointerlineskip}
     \upbrall\crcr\noalign{\kern1pt
   }}}}\limits}

\newcommand{\sg}{\sigma}

\newcommand{\oa}{\omega}
\newcommand{\al}{\alpha}
\newcommand{\va}{\varpi}
\newcommand{\nn}{\nonumber}

\newcommand{\bA}{\boldsymbol{A}}
\newcommand{\bB}{\boldsymbol{B}}
\newcommand{\bC}{\boldsymbol{C}}
\newcommand{\bI}{\boldsymbol{I}}
\newcommand{\bK}{\boldsymbol{K}}
\newcommand{\bL}{\boldsymbol{L}}
\newcommand{\bM}{\boldsymbol{M}}

\newcommand{\br}{\boldsymbol{r}}
\newcommand{\bs}{\boldsymbol{s}}
\newcommand{\bu}{\boldsymbol{u}}

\newcommand{\bT}{\boldsymbol{T}}
\newcommand{\bF}{\boldsymbol{F}}
\newcommand{\bG}{\boldsymbol{G}}
\newcommand{\bH}{\boldsymbol{H}}

\newcommand{\Ga}{\boldsymbol{\Gamma}}
\newcommand{\boa}{\boldsymbol{\omega}}
\newcommand{\Lb}{\boldsymbol{\Lambda}}

\newcommand{\st}{\hbox{\tiny\it{T}}}

\begin{document}

\title{{Kadomtsev-Petviashvili system and reduction: generalized Cauchy matrix approach}}

\author{Song-lin Zhao\footnote{Corresponding author. Email: songlinzhao1984@gmail.com}~,~~Shou-feng Shen,~~Wei Feng\\
\\
{\small \it Department of Applied Mathematics, Zhejiang University of Technology, Hangzhou 310023, P.R. China}
}

\date{}
\maketitle

\begin{abstract}

By the Sylvester equation $\bL\bM-\bM\bK=\br\bs^{\st}$ together with an evolution equation set of $\br$ and $\bs$,
generalized Cauchy matrix approach is established to investigate exact solutions for Kadomtsev-Petviashvili system, including Kadomtsev-Petviashvili equation, modified Kadomtsev-Petviashvili equation and Schwarzian Kadomtsev-Petviashvili equation.
The matrix $\bM$ provides $\tau$-function by $\tau=|\bI+\bM\bC|$. With the help of some recurrence relations,
the reduction to Korteweg-de Vries system, Boussinesq system and extended Boussinesq system are also discussed.



\end{abstract}

\maketitle

%

\section{Introduction}
\label{sec-1}

As one of the most well-known matrix equations, Sylvester equation
\begin{equation}
\bA\bM-\bM\bB=\bC
\label{SE}
\end{equation}
has attracted lots of attention and substantial
progress of this equation has been made \cite{S-1884,M-book-1987}.
Equation \eqref{SE}, containing the Lyapunov equation as a special
case, plays an important role in many areas of applied mathematics, such as control theory,
signal processing, filtering, model reduction, image restoration.
The solvability to the Sylvester equation
\eqref{SE} has been considered in many references \cite{SE-sol,BR-BLMS-1997}. In Ref. \cite{BR-BLMS-1997}, Bhatia and Rosenthal
investigated many interesting and important applications
related to the solutions of equation \eqref{SE}, including similarity, commutativity, hyperinvariant
subspaces, spectral operators, differential equations,  and so on.

With the help of the Sylvester equation,  many methods have been introduced to solve the integrable systems, such as
operator method \cite{AC-1996-JMP,S-PD-1998}, bidifferential calculus approach
\cite{DM-DCDS-2009,HL-PLA-2001}, method based on
Gel'fand-Levitan-Marchenko equation \cite{AM-IP-2006}
and Cauchy matrix approach \cite{NAH-2009-JPA}. The Cauchy matrix approach
\cite{NAH-2009-JPA} was firstly proposed by Nijhoff and his collaborators
to investigate the soliton solutions to lattice Korteweg-de Vries (KdV)-type equations and
Adler-Bobenko-Suris (ABS) lattice. In this method,
a Cauchy-type matrix $\bM=(M_{i,j})_{N\times N},~M_{i,j}=\frac{r_i s_j}{k_i+k_j}$ was introduced,
which satisfies a Sylvester equation
\begin{equation}
\bK\bM+\bM \bK = \br\, \bs^{\st},
\label{SE-1}
\end{equation}
where $\bK=\mathrm{Diag}(k_1,k_2,\cdots,k_N)$; $\br=(r_1,r_2,\cdots,r_N)^{\st}$ with $r_i=\bigl(\frac{a-k_i}{a+k_i}\bigr)^n\bigl(\frac{b-k_i}{b+k_i}\bigr)^m r^{(0)}_i$ is a
known plain wave factor column vector and $\bs=(s_1,s_2,\cdots,s_N)^{\st}$ is a constant column vector. By defining
scalar function $\S{i,j}=\bs^{\st}\,\bK^j(\bI+\bM)^{-1}\bK^i\br$ and considering its
dynamical properties, Nijhoff {\it et al.} constructed the soliton solutions for
lattice KdV system, including lattice KdV equation, lattice modified KdV equation,
lattice Schwarzian KdV equation and Nijhoff-Quispel-Capel (NQC) equation. Furthermore, with
the relation between NQC equation and Q3$_0$ equation and the degeneration relations
among equations Q3, Q2, Q1, H3, H2 and H1, the soliton solutions
for ABS lattice were considered.
By using a general Sylvester equation \cite{N-2004-math}:
\begin{equation}
\bL\bM+\bM\bK=\br\, \bs^{\st},
\label{SE-11}
\end{equation}
where $\bK,~\bL$ are diagonal constant matrices and $\br,\,\bs$ are known column vectors,
Nijhoff also constructed the lattice Kadomtsev-Petviashvili (KP) system and obtained their soliton solutions,
including lattice KP equation, lattice modified KP equation, lattice Schwarzian KP equation
and KP-type NQC equation. Motivated by these two works,
an alternative method was developed, called ``generalized Cauchy matrix approach'',
which has been used to derive several kinds of solutions for autonomous discrete integrable system
\cite{ZZ-SAM-2013,ZZS-2012,FZZ-2012,FZ-2013} as well as non-autonomous discrete integrable system
\cite{ZSF-2014}.
Recently, this method was also utilized to discuss KdV system and sine-Gordon
equation \cite{XZZ-2014}, which can be viewed as an
application in continue integrable system.

In present paper, we will explore continue KP
system by applying the generalized Cauchy matrix approach. The method uses a Sylvester equation
similar to \eqref{SE-11} associated with
some evolution relations on vectors $\br$ and $\bs$, which are called determining equation set (DES).
Then by discussing the dynamical properties of scalar functions $\S{i,j}$,
we derive KP-type equations, including KP equation, modified KP equation and Schwarzian KP equation. Besides,
the connection with $\S{i,j}$ and $\tau$-function is also discussed. In view of
some special forms of matrices $\bK$ and $\bL$, we consider the reduction of KP system.
As a consequence, KdV system, Boussinesq (BSQ) system and extended BSQ system are constructed.

The paper is organized as follows.
In Sec.\ref{sec-2}, we first set up the DES.
In addition, we introduce scalar function $\S{i,j}$ and
discuss its dynamical properties.
Next, for different values of $i$ and $j$, we construct KP-type equations.
In Sec.\ref{sec-3}, exact solutions
to the DES are obtained.
Sec.\ref{sec-4} is devoted to the discussion on the reduction of KP system. Finally,
some conclusions are made in Sec. 5.

\section{The DES and KP system}\label{sec-2}

In this section, the generalized Cauchy matrix approach for the
continue KP system will be established. Firstly, we set up the DES.
Next, scalar function $\S{i,j}$ will be introduced and its
dynamical properties will be discussed. Then due to
the values of $i$ and $j$, KP equation, modified KP equation and Schwarzian KP equation
will be constructed. $\tau$-function will be discussed in the last subsection.

In generalized Cauchy matrix approach, the following Proposition \cite{S-1884} is always needed.
\begin{Proposition} \label{solv}
Let us denote the eigenvalue sets of matrices $\bA$ and $\bB$ by $\mathcal{E}(\bA)$ and $\mathcal{E}(\bB)$, respectively.
For the known matrices $\bA, \bB$ and $\bC$, the Sylvester equation \eqref{SE}
has a unique solution $\bM$ if and only if $\mathcal{E}(\bA)\bigcap \mathcal{E}(\bB)=\varnothing$.
\end{Proposition}
With some more conditions on
$\mathcal{E}(\bA)$ and $\mathcal{E}(\bB)$, solution $\bM$ of \eqref{SE} can be expressed
via series or integration \cite{BR-BLMS-1997} (see also Ref. \cite{XZZ-2014}).

\subsection{The DES} \label{subsec-2.1}

To proceed, we consider a Sylvester equation in form of
\begin{equation}
\bL \bM-\bM\bK=\br\, \bs^{\st},
\label{SE-2}
\end{equation}
where $\bL \in \mathbb{C}_{N\times N}$, $\bK \in \mathbb{C}_{N'\times N'}$, $\bM \in \mathbb{C}_{N\times N'}$,
$\br=(r_1,r_2,\cdots,r_N)^{\st}$ and $\bs=(s_1,s_2,\cdots,s_{N'})^{\st}$.
Equation \eqref{SE-2} is solvable and has unique solution $\bM$ when $\mathcal{E}(\bK)\bigcap \mathcal{E}(\bL)=\varnothing$.
In the rest part of this section, we assume that $\bK$ and $\bL$ satisfy such condition and $0 \notin \mathcal{E}(\bK),~\mathcal{E}(\bL)$,
i.e., $\bK$ and $\bL$ are invertible matrices. In discrete system, the Sylvester equation \eqref{SE-2}
has been used to construct the lattice KP system and
their various solutions \cite{FZ-2013}, where $\br$ and $\bs$ satisfy some
discrete evolution equations. In order to investigate the application of \eqref{SE-2} in continue KP system,
we suppose the following evolution equation set
\begin{subequations}
\begin{align}
& \br_x=\bL \br,~~\bs_x=-\bK^{\st} \bs, \label{evo-rs-x} \\
& \br_y=-\bL^2 \br,~~\bs_y=(\bK^{\st})^2 \bs, \label{evo-rs-y} \\
& \br_t=4\bL^3 \br,~~\bs_t=-4(\bK^{\st})^3\bs, \label{evo-rs-t}
\end{align}
\label{evo-rs}
\end{subequations}
where $\br, \bs$ and $\bM$ are functions of $(x,y,t)$ while $\bK$ and $\bL$ are non-trivial constant matrices.
Equations \eqref{SE-2} and \eqref{evo-rs} are the so-called DES,
which plays a basic role in the generalized Cauchy matrix approach \cite{ZZ-SAM-2013}.
Among the DES, evolution equations \eqref{evo-rs}
are used to determine plain wave factor vectors $\br$ and $\bs$, and Sylvester equation \eqref{SE-2} is
used to define matrix $\bM$.

We now discuss the dynamical properties of matrix $\bM$, i.e., the evolution relations
of $\bM$ w.r.t. independent variables. The derivative of the Sylvester equation \eqref{SE-2}
w.r.t. $x$ together with \eqref{evo-rs-x} yields
\[
\bL \bM_x-\bM_x\bK=\br_x \,\bs^{\st}+ \br \bs^{\st}_x=\bL \br \bs^{\st}- \br \bs^{\st} \bK,
\]
which gives rise to the relation
\begin{equation}
\bM_x=\br \bs^{\st},
\label{evo-Mx}
\end{equation}
in the light of Proposition \ref{solv}.

The $y$-derivative of equation \eqref{SE-2} leads to
\begin{align}
\bL \bM_y-\bM_y \bK & = \br_y \,\bs^{\st}+ \br \bs^{\st}_y \nn \\
& =-\bL^2 \br \bs^{\st}+ \br \bs^{\st} \bK^2 \nn \\
& =\bL(-\bL^2\bM+ \bM\bK^2)-(-\bL^2\bM+ \bM\bK^2)\bK, \label{My-eq}
\end{align}
where in the last step the term $\br \bs^{\st}$ is replaced by $\bL \bM-\bM\bK$.
Then we get
\begin{equation}
\bM_y=-\bL^2\bM+ \bM\bK^2,
\label{evo-My}
\end{equation}
which can be rewritten as
\begin{align}
\bM_y=-\br\bs^{\st}\bK-\bL\br\bs^{\st}. \label{evo-My2}
\end{align}
Analogous to the earlier analysis, we deduce that the time evolution of $\bM$ is of form
\begin{align}
\bM_t & =4(\bL^3 \bM- \bM\bK^3) \nn \\
      & =4(\bL^2\br\bs^{\st}+\bL\br\bs^{\st}\bK+\br\bs^{\st}\bK^2).
\label{evo-Mt}
\end{align}
\eqref{evo-Mx}, \eqref{evo-My} and \eqref{evo-Mt} encode all the information on the dynamics of the matrix $\bM$, w.r.t. the
independent variables $x$, $y$ and $t$, in addition to \eqref{SE-2} which can be thought as the defining property of $\bM$.

\subsection{Objects $\S{i,j}$} \label{subsec-2.2}

\subsubsection{The definition of $\S{i,j}$} \label{subsubsec-2.2.1}

By the Sylvester equation \eqref{SE-2}, we now introduce a scalar function
\begin{equation}
S^{(i,j)}=\bs^{\st}\,\bK^j\bC(\bI+\bM\bC)^{-1}\bL^i\br, ~~i,j\in \mathbb{Z},
\label{Sij}
\end{equation}
where $\bI$ is the $N$th-order unit matrix; $\bC \in \mathbb{C}_{N'\times N}$ is an
arbitrary constant matrix, such that the product $\bM\bC$ is a square $N\times N$ matrix.
Similar to Ref. \cite{XZZ-2014}, here $\S{i,j}$ is also called a master function since
it will be used to generate integrable equations.
For convenience, we introduce an auxiliary vector function
\begin{equation}
\bu^{(i)}=(\bI+\bM\bC)^{-1}\bL^i\br, ~~i\in \mathbb{Z}.
\label{ui}
\end{equation}
Then $S^{(i,j)}$ defined in \eqref{Sij} can be simplified to
\begin{equation}
S^{(i,j)}=\bs^{\st}\,\bK^j\bC\bu^{(i)}, ~~i,j\in \mathbb{Z}.
\label{Sij-ui}
\end{equation}
It is noteworthy that $S^{(i,j)}$ is not symmetric w.r.t. the interchange of
the parameters $i$ and $j$, i.e. $S^{(i,j)}\neq S^{(j,i)}$.

\subsubsection{Evolution of $\S{i,j}$} \label{subsubsec-2.2.2}

To begin, we consider the dynamical properties of the vector function $\bu^{(i)}$ defined by \eqref{ui}.
It follows from \eqref{ui} that equation
\begin{equation}
(\bI+\bM\bC)\bu^{(i)}=\bL^i\br
\label{ui1}
\end{equation}
holds identically.
Substituting \eqref{evo-rs-x} and \eqref{evo-Mx} into the $x$-derivative of equation \eqref{ui1} yields
\[
(\bI+\bM\bC)\bu^{(i)}_x=\bL^{i+1}\br-\br\bs^{\st}\bC\bu^{(i)},
\]
which implies
\begin{equation}
\bu^{(i)}_x=\bu^{(i+1)}-S^{(i,0)}\bu^{(0)},
\label{evo-uix}
\end{equation}
where relation \eqref{Sij-ui} has been used.
After a similar analysis as aforementioned,
we arrive at the evolution of $\bu^{(i)}$ in $y$, $t$-directions
\begin{subequations}
\begin{align}
& \bu^{(i)}_y =-\bu^{(i+2)}+S^{(i,0)}\bu^{(1)}+S^{(i,1)}\bu^{(0)}, \label{evo-uiy} \\
& \bu^{(i)}_t =4(\bu^{(i+3)}-S^{(i,2)}\bu^{(0)}-S^{(i,1)}\bu^{(1)}-S^{(i,0)}\bu^{(2)}). \label{evo-uit}
\end{align}
\label{evo-uiyt}
\end{subequations}
Multiplying \eqref{evo-uix} and \eqref{evo-uiyt}
from the left by the row vector $\bs^{\st}\bK^j\bC$ and noting that the equation \eqref{evo-rs}
and the connection \eqref{Sij-ui} between $\bu^{(i)}$ and $\S{i,j}$,
we have the evolution relations of $S^{(i,j)}$:
\begin{subequations}
\begin{align}
& S^{(i,j)}_{x}=S^{(i+1,j)}-S^{(i,j+1)}-S^{(i,0)}S^{(0,j)}, \label{evo-Sijx} \\
& S^{(i,j)}_{y}=-S^{(i+2,j)}+S^{(i,j+2)}+S^{(i,1)}S^{(0,j)}+S^{(i,0)}S^{(1,j)}, \label{evo-Sijy} \\
& S^{(i,j)}_t=4(S^{(i+3,j)}-S^{(i,j+3)}-S^{(i,0)}S^{(2,j)}-S^{(i,1)}S^{(1,j)}-S^{(i,2)}S^{(0,j)}). \label{evo-Sijt}
\end{align}
It is readily to obtain
some higher-order derivatives of $S^{(i,j)}$ w.r.t. independent variables from above relations
by iterate calculation.
Here we just present the expressions of $S^{(i,j)}_{xx}$, $S^{(i,j)}_{xxx}$ and $S^{(i,j)}_{xy}$ as follows:
\begin{align}
S^{(i,j)}_{xx}=& S^{(i+2,j)}+S^{(i,j+2)}-2S^{(i+1,j+1)}-2S^{(i+1,0)}S^{(0,j)}+2S^{(i,0)}S^{(0,j+1)}\nn\\
& -S^{(i,0)}S^{(1,j)}+S^{(i,1)}S^{(0,j)}+2S^{(0,0)}S^{(i,0)}S^{(0,j)}, \label{evo-Sijxx}\\
S^{(i,j)}_{xxx}=& S^{(i+3,j)}-S^{(i,j+3)}-3S^{(i+2,j+1)}+3S^{(i+1,j+2)}-3S^{(i+2,0)}S^{(0,j)}\nn\\
&  -3S^{(i,0)}S^{(0,j+2)}+6S^{(i+1,0)}S^{(0,j+1)}-3S^{(i+1,0)}S^{(1,j)}-3S^{(i,1)}S^{(0,j+1)}\nn\\
&  -S^{(i,2)}S^{(0,j)}-S^{(i,0)}S^{(2,j)}+3S^{(i+1,1)}S^{(0,j)}+3S^{(i,0)}S^{(1,j+1)}\nn\\
&  +6S^{(i+1,0)}S^{(0,0)}S^{(0,j)}-6S^{(i,0)}S^{(0,0)}S^{(0,j+1)}+2S^{(i,1)}S^{(1,j)}\nn\\
&  +3S^{(i,0)}S^{(0,0)}S^{(1,j)}+3S^{(i,0)}S^{(1,0)}S^{(0,j)}-3S^{(i,0)}S^{(0,1)}S^{(0,j)}\nn\\
&   -3S^{(i,1)}S^{(0,0)}S^{(0,j)}-6S^{(i,0)}{S^{(0,0)}}^2 S^{(0,j)}, \label{evo-Sijxxx} \\
S^{(i,j)}_{xy}=& -S^{(i+3,j)}-S^{(i,j+3)}+S^{(i+2,j+1)}+S^{(i+1,j+2)}+S^{(i+2,0)}S^{(0,j)}\nn\\
&  -S^{(i,0)}S^{(0,j+2)}-S^{(i,1)}S^{(0,j+1)}+S^{(i+1,0)}S^{(1,j)}-S^{(i,2)}S^{(0,j)}+S^{(i,0)}S^{(2,j)}\nn\\
&  +S^{(i+1,1)}S^{(0,j)}-S^{(i,0)}S^{(1,j+1)}
 -S^{(i,1)}S^{(0,0)}S^{(0,j)}-S^{(i,0)}S^{(0,0)}S^{(1,j)}\nn\\
& -S^{(i,0)}S^{(0,1)}S^{(0,j)}-S^{(i,0)}S^{(1,0)}S^{(0,j)}. \label{evo-Sijxy}
\end{align}
The relation \eqref{evo-Sijx} implies that the following indentities
\begin{align}
& S^{(i,j+2)}=-S^{(i,j+1)}_{x}+S^{(i+1,j+1)}-S^{(0,j+1)}S^{(i,0)}, \label{Sij+2} \\
& S^{(i+2,j)}=S^{(i+1,j)}_{x}+S^{(i+1,j+1)}+S^{(0,j)}S^{(i+1,0)} \label{Si+2j}
\end{align}
hold.
Thus the subtraction of \eqref{Sij+2} from \eqref{Si+2j} leads to
\begin{align}
S^{(i+2,j)}-S^{(i,j+2)}=\partial_x(S^{(i,j+1)}+S^{(i+1,j)})+S^{(0,j+1)}S^{(i,0)}+S^{(0,j)}S^{(i+1,0)}.
\label{Si+2j-Sij+2}
\end{align}
Plugging \eqref{Si+2j-Sij+2} into \eqref{evo-Sijy} and utilizing
\begin{align}
& S^{(0,j+1)}=-S^{(0,j)}_{x}+S^{(1,j)}-S^{(0,j)}S^{(0,0)}, \label{Sij+1} \\
& S^{(i+1,0)}=S^{(i,0)}_{x}+S^{(i,1)}+S^{(i,0)}S^{(0,0)}, \label{Si+1j}
\end{align}
we finally arrive at
\begin{eqnarray}
&& \partial^{-1} S_{y}^{(i,j)} =-(S^{(i,j+1)}+S^{(i+1,j)})+\partial^{-1}(S^{(i,0)}S^{(0,j)}_{x}-S_x^{(i,0)}S^{(0,j)}), \label{pSijy} \\
&& \partial^{-1} S_{yy}^{(i,j)} =S^{(i+3,j)}-S^{(i,j+3)}+S^{(i+2,j+1)}-S^{(i+1,j+2)}-S^{(0,j)}S^{(i+1,1)} \nn  \\
&&~~~~~~~~~~~~~~ -S^{(0,j+1)}S^{(i,1)}-S^{(1,j)}S^{(i+1,0)}-S^{(1,j+1)}S^{(i,0)} \nn  \\
&&~~~~~~~~~~~~~~+\partial_y\partial^{-1}(S^{(i,0)}S^{(0,j)}_{x}-S_x^{(i,0)}S^{(0,j)}), \label{pSijyy}
\end{eqnarray}
\label{deri-var-Sij}
\end{subequations}
where $\partial^{-1}=\frac{1}{2}(\int_{-\infty}^x-\int_{x}^\infty)$
and relation \eqref{pSijyy} is derived from \eqref{pSijy} by taking $y$-derivative. All
the relations in \eqref{deri-var-Sij} can be viewed as semi-discrete equations when the
parameters $i$ and $j$ are recognized as discrete independent variables.

\subsubsection{Invariance of $\S{i,j}$ }\label{subsubsec-2.2.3}

Now suppose that matrices $\bK_1$ and $\bL_1$ are similar to $\bK$ and $\bL$,
respectively, under the transform matrices $\bT_1$
and $\bT_2$, i.e.
\begin{subequations}
\label{trans-sim}
\begin{equation}
\bK_1=\bT_1 \bK \bT_1^{-1},~~\bL_1=\bT_2 \bL \bT_2^{-1}.
\label{KL1-KL}
\end{equation}
We denote
\begin{equation}
\bM_1=\bT_2 \bM \bT_1^{-1},~~\bC_1=\bT_1 \bC \bT_2^{-1},~~\br_1=\bT_2 \br,~~\bs_1^{\st}=\bs^{\st} \bT_1^{-1}.
\label{Mrs-1}
\end{equation}
\end{subequations}
Then by direct substituting, DES \eqref{SE-2} and \eqref{evo-rs} yield
\begin{subequations}
\begin{align}
& \bL_1\bM_1-\bM_1\bK_1= \br_1 \,\bs^{\st}_1,\label{SE-co-inv} \\
& \br_{1_x}=\bL_1 \br_1,~~\bs_{1_x}=-\bK_1^{\st} \bs_1, \label{evo-rs-x-inv} \\
& \br_{1_y}=-\bL_1^2 \br_1,~~\bs_{1_y}=(\bK_1^{\st})^2 \bs_1, \label{evo-rs-y-inv} \\
& \br_{1_t}=4\bL_1^3 \br_1,~~\bs_{1_t}=-4(\bK_1^{\st})^3\bs_1, \label{evo-rs-t-inv}
\end{align}
and
\begin{align}
S^{(i,j)} = \bs^{\st}\, \bK^j\bC(\bI+ \bM\bC)^{-1} \bL^i \br
= \bs_1^{\st}\, \bK_1^j\bC_1(\bI+ \bM_1\bC_1)^{-1}\bK_1^i \br_1.
\end{align}
\end{subequations}
It is shown that scalar function $\S{i,j}$ is invariant under the transformation \eqref{trans-sim}.

\subsubsection{Some identities with $\S{i,j}$} \label{subsubsec-2.2.4}

With some special relations between matrices $\bL$ and $\bK$, one can derive several
important equalities for the master function $S^{(i,j)}$, which will be used in Sec. \ref{sec-4}.
Here we suppose the orders of matrices $\bK$ and $\bL$ are the same, i.e., $N'=N$ and
the constant matrix $\bC$ in \eqref{Sij} is the $N$th-order unit matrix.

\begin{Proposition}\label{prop-KdV}

Assuming $\bK=-\bL$, then for the master function $S^{(i,j)}$ with $\bM, \bL, \br,\bs$
satisfying the Sylvester equation \eqref{SE-2}, we have the following relation
\begin{equation}
\S{i,j+2k}=\S{i+2k,j}-\sum^{2k-1}_{l=0}\S{2k-1-l,j}\S{i,l},~~~(k=1,2,\cdots).
\label{Sij-2k+}
\end{equation}
In particular, when $k=1$ we have
\begin{equation}
S^{(i,j+2)}=S^{(i+2,j)}-S^{(i,0)}S^{(1,j)}-S^{(i,1)}S^{(0,j)}.
\label{Sij-2k=1+}
\end{equation}
\end{Proposition}
The proof of Proposition \ref{prop-KdV} can be referred to Ref. \cite{XZZ-2014}.
The relation \eqref{Sij-2k=1+} firstly appeared in discrete case \cite{NAH-2009-JPA} plays a
crucial role in the construction of the continue integrable systems by using generalized Cauchy matrix
approach (see Ref. \cite{XZZ-2014}).
It is worth noting that the master function $\S{i,j}$ in Proposition \ref{prop-KdV} is of form
\[\S{i,j}=(-1)^j\bs^{\st}\,\bL^j(\bI+\bM)^{-1}\bL^i\br,\]
which has symmetric property $\S{i,j}=\S{j,i}$ when $j-i$ is an
even number and antisymmetric property $\S{i,j}=-\S{j,i}$ when $j-i$ is an
odd number \cite{ZZ-SAM-2013,XZZ-2014}.

Besides the above case, we have the following two results, where parameter $\oa$
satisfies $\oa^2+\oa+1=0$.
\begin{Proposition}\label{prop-BSQ-1}
Assuming $\bK=\oa\bL$, then for the master function $S^{(i,j)}$  with $\bM, \bL, \br,\bs$
satisfying the Sylvester equation \eqref{SE-2}, we have the following relation
\begin{equation}
\S{i,j+3k}=\S{i+3k,j}-\sum^{3k-1}_{l=0}\S{3k-1-l,j}\S{i,l},~~~(k=1,2,\cdots).
\label{Sij-3k+}
\end{equation}
In particular, when $k=1$ we have
\begin{equation}
S^{(i,j+3)}=S^{(i+3,j)}-S^{(i,2)}S^{(0,j)}-S^{(i,1)}S^{(1,j)}-S^{(i,0)}S^{(2,j)}.
\label{Sij-3k=1+}
\end{equation}
\end{Proposition}
\begin{Proposition}\label{prop-BSQ-2}
Assume $\bK=\mbox{Diag}(\oa\bK_1,\oa^2\bK_2)$, $\bL=\mbox{Diag}(\bK_1,\bK_2)$,
where $\bK_i \in \mathbb{C}_{N_i\times N_i}$ with $N_1+N_2=N$. Then for the master function $S^{(i,j)}$  with $\bM, \bL, \bK, \br,\bs$
satisfying the Sylvester equation \eqref{SE-2}, we have the relations \eqref{Sij-3k+} and \eqref{Sij-3k=1+}.
\end{Proposition}
The proofs of Proposition \ref{prop-BSQ-1} and Proposition \ref{prop-BSQ-2} are similar to the one for Proposition \ref{prop-KdV},
which are omitted here.
Furthermore, for the roots $\boa_1(\bK),~\boa_2(\bK)$ and $\boa_3(\bK)=\bK$ of the matrix algebraic relation
\begin{eqnarray}
\label{eq:g}
G_{3}(\boa,\bK):=g(\boa)-g(\bK)=0,~{\rm where}~g(\bK)=\sum_{j=1}^{3}\al_j\bK^{j},
\end{eqnarray}
with $\al_i \in \mathbb{C}~(i=1,2)$ and $\al_{3}=1,$
we have the following result.
\begin{Proposition}\label{prop-exBSQ}
Assume $\bK=\mbox{Diag}(\boa_1(\bK_1),\boa_2(\bK_2))$, $\bL=\mbox{Diag}(\bK_1,\bK_2)$,
where $\bK_i \in \mathbb{C}_{N_i\times N_i}$ with $N_1+N_2=N$. Then for the master function $S^{(i,j)}$  with $\bM, \bL, \bK, \br,\bs$
satisfying the Sylvester equation \eqref{SE-2}, we have the relation
\begin{eqnarray}
&& S^{(i+3,j)}-S^{(i,j+3)}-S^{(i,2)}S^{(0,j)}-S^{(i,1)}S^{(1,j)}-S^{(i,0)}S^{(2,j)} \nn \\
&&~~~~= \al_2(S^{(i,j+2)}-S^{(i+2,j)}+S^{(i,0)}S^{(1,j)}+S^{(i,1)}S^{(0,j)}) \nn \\
&& ~~~~~~~+\al_1(S^{(i,j+1)}-S^{(i+1,j)}+S^{(i,0)}S^{(0,j)}).
\label{Sij-3alk=1+}
\end{eqnarray}
\end{Proposition}

For more details of Proposition \ref{prop-exBSQ}, one can see Appendix \ref{A:G3}.

\subsection{The KP system}\label{subsec-2.3}

From the evolution relation \eqref{deri-var-Sij}, various KP-type equations can be
constructed for special values of the parameters $i$ and $j$, including KP equation, modified KP equation and Schwarzian KP equation.

\subsubsection{The KP equation}

To derive the KP equation, we take $i=j=0$ in \eqref{deri-var-Sij} and denote $u=\S{0,0}$.
In this case, some evolution relations in \eqref{deri-var-Sij} give rise to
\begin{subequations}
\begin{align}
& u_{x}=-S^{(0,1)}+S^{(1,0)}-u^{2}, \label{ux} \\
& u_{t}=4(S^{(3,0)}-S^{(0,3)}-uS^{(2,0)}-uS^{(0,2)}-S^{(1,0)}S^{(0,1)}),\\
& u_{xxx}=-S^{(0,3)}+S^{(3,0)}+3(S^{(1,2)}-S^{(2,1)})-4u(S^{(2,0)}+S^{(0,2)})+8S^{(1,0)}S^{(0,1)} \nn \\
&  ~~~~~~~~~-3(S^{(1,0)^2}+S^{(0,1)^2})+6uS^{(1,1)}+12u^2(S^{(1,0)}-S^{(0,1)})-6u^4, \\
& \partial^{-1}u_{yy}=S^{(3,0)}-S^{(0,3)}+S^{(2,1)}-S^{(1,2)}-2uS^{(1,1)}-S^{(0,1)^2}-S^{(1,0)^2}.
\end{align}
\label{u-deri}
\end{subequations}
From \eqref{u-deri}, one can easily find the following potential KP equation
\begin{eqnarray}
u_{t}-u_{xxx}-6u_{x}^{2}-3\partial^{-1}u_{yy}=0. \label{pKP}
\end{eqnarray}
By transformation $\va=2u_x$, equation \eqref{pKP} is transformed to the KP equation
\begin{eqnarray}
\va_{t}-\va_{xxx}-6\va \va_{x}-3\partial^{-1}\va_{yy}=0, \label{KP}
\end{eqnarray}
which has solution
\begin{equation}
\va=2(\bs^{\st}\bC(\bI+\bM\bC)^{-1}\br)_x,
\end{equation}
where matrix $\bM$ and vectors $\bs$, $\br$ satisfy
DES \eqref{SE-2} and \eqref{evo-rs}.

\subsubsection{The modified KP equation}

To derive the modified KP equation, we consider the following two cases:
\begin{align}
& i=0 \text{ and } j=-1;
\label{mkp-case-1}\\
& i=-1 \text{ and } j=0.
\label{mkp-case-2}
\end{align}

For case \eqref{mkp-case-1}, a new variable $v=S^{(0,-1)}+1$ is introduced.
Then some evolution relations in \eqref{deri-var-Sij} become
\begin{subequations}
\begin{align}
&v_{x}= S^{(1,-1)}-uv, \label{vx} \\
&v_{t}=4(S^{(3,-1)}-vS^{(0,2)}-S^{(1,-1)}S^{(0,1)}-S^{(2,-1)}u),\\
&v_{xx}= vS^{(0,1)}+S^{(2,-1)}-2vS^{(1,0)}+2u^{2}v-uS^{(1,-1)}, \label{vxx} \\
&v_{xxx}= S^{(3,-1)}-vS^{(0,2)}+3vS^{(1,1)}-3vS^{(2,0)}-6uvS^{(0,1)}+9uvS^{(1,0)}\nn \\
&~~~~~~~~-6u^{3}v-uS^{(2,-1)}-3S^{(1,0)}S^{(1,-1)}+2S^{(0,1)}S^{(1,-1)}+3u^{2}S^{(1,-1)}, \\
&v_{y}= vS^{(0,1)}-S^{(2,-1)}+uS^{(1,-1)}, \label{vy} \\
& v_{xy}= -S^{(3,-1)}-vS^{(0,2)}+vS^{(1,1)}+vS^{(2,0)}-2uvS^{(0,1)}-uvS^{(1,0)}\nn \\
&~~~~~~~~ +uS^{(2,-1)}+S^{(1,0)}S^{(1,-1)}-u^{2}S^{(1,-1)}.
\end{align}
\label{v-deri}
\end{subequations}
After a straightforward computation, we obtain equation
\begin{align}
v_{t}-v_{xxx}-6u_{x}v_{x}+6vu_y+3v_{xy}=0, \label{v-u-eq}
\end{align}
where $u_{x}$ is given by \eqref{ux} and
\begin{align}
u_{y}=-S^{(2,0)}+S^{(0,2)}+S^{(0,1)}u+uS^{(1,0)}. \label{uy}
\end{align}
Besides, noting that \eqref{vxx}, \eqref{vy} and \eqref{ux}, we also have
\begin{align}
v_{xx}=-2vu_x-v_y. \label{vxx-vy}
\end{align}
Substituting \eqref{vxx-vy} into \eqref{v-u-eq} and with direct computation, we get the potential modified KP equation
\begin{align}
v_{t}-v_{xxx}+3\frac{v_xv_{xx}}{v}+6v\partial^{-1}\bigg(\bigg(\frac{v_x}{v}\bigg)_x\frac{v_y}{v}\bigg)
-3v\partial^{-1}\bigg(\frac{v_{yy}v-v_y^2}{v^2}\bigg)=0. \label{v-mKP}
\end{align}

By transformation
$\mu= \partial_x \ln v,$
equation \eqref{v-mKP} is transformed into modified KP equation
\begin{eqnarray}
\mu_{t}-\mu_{xxx}+6\mu^2\mu_x+6\mu_x\partial_x^{-1}\mu_y-3\partial^{-1}\mu_{yy}=0, \label{mKP}
\end{eqnarray}
which possesses solution
\begin{equation}
\mu=\partial_x \ln (1+\bs^{\st}\bK^{-1}\bC(\bI+\bM\bC)^{-1}\br),
\end{equation}
where matrices $\bM,~\bK$ and vectors $\bs$, $\br$ satisfy
DES \eqref{SE-2} and \eqref{evo-rs}.

Noting that $\va=2u_x$ and $\mu= \partial_x \ln v$, relation \eqref{vxx-vy} implies
\begin{eqnarray}
-\va=\mu_x+\mu^2+\partial^{-1}\mu_y, \label{MT-KP-mKP}
\end{eqnarray}
which is the Miura transformation between modified KP equation \eqref{mKP} and KP equation \eqref{KP}.

For case \eqref{mkp-case-2}, we consider a new variable
\begin{equation}
w=1-S^{(-1,0)}, \label{w-def}
\end{equation}
whose various derivatives can be derived directly from \eqref{deri-var-Sij} with $ i=-1 \text{ and } j=0$:
\begin{subequations}
\begin{align}
&w_{x}=S^{(-1,1)}-uw, \label{wx} \\
&w_{t}=4(S^{(-1,3)}-wS^{(2,0)}+S^{(-1,1)}S^{(1,0)}+S^{(-1,2)}u),\\
&w_{xx}=-wS^{(1,0)}-S^{(-1,2)}+2wS^{(0,1)}+2u^{2}w-uS^{(-1,1)}, \label{wxx} \\
&w_{xxx}=S^{(-1,3)}-wS^{(2,0)}-3wS^{(0,2)}+3wS^{(1,1)}+6uwS^{(1,0)}-9uwS^{(0,1)}\nn \\
&~~~~~~~~~ -6u^{3}w+uS^{(-1,2)}+3S^{(0,1)}S^{(-1,1)}-2S^{(1,0)}S^{(-1,1)}+3u^{2}S^{(-1,1)}, \\
&w_{y}= -S^{(-1,2)}-uS^{(-1,1)}+wS^{(1,0)}, \label{wy} \\
&w_{xy}= S^{(-1,3)}-wS^{(0,2)}-wS^{(1,1)}+wS^{(2,0)}-2uwS^{(1,0)}-uwS^{(0,1)}\nn \\
&~~~~~~~~+uS^{(1,-2)}+S^{(0,1)}S^{(-1,1)}+u^{2}S^{(-1,1)}.
\end{align}
\end{subequations}
A straightforward calculation yields equations
\begin{subequations}
\begin{align}
& w_{t}-w_{xxx}-6u_{x}w_{x}-6wu_y-3w_{xy}=0, \label{w-u-eq} \\
& w_{xx}=-2wu_x+w_y, \label{wxx}
\end{align}
\end{subequations}
where $u_x$ and $u_y$ are defined by \eqref{ux} and \eqref{uy}. Similarly, taking \eqref{wxx}
into \eqref{w-u-eq}, we derive one more potential modified KP equation
\begin{align}
w_{t}-w_{xxx}+3\frac{w_xw_{xx}}{w}-6w\partial^{-1}\bigg(\bigg(\frac{w_x}{w}\bigg)_x\frac{w_y}{w}\bigg)
-3w\partial^{-1}\bigg(\frac{w_{yy}w-w_y^2}{w^2}\bigg)=0. \label{w-mKP}
\end{align}
In terms of transformation
$\nu= \partial_x \ln w$,
we get the corresponding modified KP equation
\begin{eqnarray}
\nu_{t}-\nu_{xxx}+6\nu^2\nu_x-6\nu_x\partial^{-1}\nu_y-3\partial^{-1}\nu_{yy}=0, \label{an-mKP}
\end{eqnarray}
whose solution is given by
\begin{equation}
\nu=\partial_x \ln (1-\bs^{\st}\bC(\bI+\bM\bC)^{-1}\bL^{-1}\br),
\end{equation}
where matrices $\bM,~\bL$ and vectors $\bs,~\br$ satisfy
DES \eqref{SE-2} and \eqref{evo-rs}.
It is easy to know that $-\mu$ also satisfies the modified KP equation \eqref{an-mKP}.

In the light of $\va=2u_x$ and $\nu= \partial_x \ln w$, relation \eqref{wxx} gives rise to
the Miura transformation between modified KP equation \eqref{an-mKP} and KP equation \eqref{KP},
i.e.
\begin{eqnarray}
-\va=\nu_x+\nu^2-\partial^{-1}\nu_y. \label{MT-KP-an-mKP}
\end{eqnarray}

\subsubsection{The Schwarzian KP equation}

Let us examine the equation related to function $\S{-1,-1}$. We introduce
\begin{eqnarray}
z=S^{(-1,-1)}+x.
\label{z-def}
\end{eqnarray}
Setting $i=j=-1$ in \eqref{deri-var-Sij} yields the following expressions
\begin{subequations}
\begin{eqnarray}
&& z_{x}=vw, \label{MT-MS} \\
&& z_{xx}=vS^{(-1,1)}+wS^{(1,-1)}-2uvw \nn \\
&&~~~~=(vw)_x , \label{zxx} \\
&& z_{xxx}=-vS^{(-1,2)}+wS^{(2,-1)}+2S^{(-1,1)}S^{(1,-1)}-3uwS^{(1,-1)}-3uvS^{(-1,1)} \nn \\
&&~~~~~~~~~-3vwS^{(1,0)}+3vwS^{(0,1)}+6u^{2}vw, \\
&& z_{t}=4(-vS^{(-1,2)}+wS^{(2,-1)}-S^{(-1,1)}S^{(1,-1)}), \\
&& z_{y}=S^{(-1,1)}v-S^{(1,-1)}w \nn \\
&&~~~=w_xv-wv_x, \label{zy}
\end{eqnarray}
\end{subequations}
where in \eqref{zxx} and \eqref{zy} we have made use of the relations \eqref{vx} and \eqref{wx}.
By a forward calculation associated with \eqref{MT-MS} and \eqref{zy}, we have
\begin{eqnarray}
&& z_{t}-z_{xxx}+\frac{3}{2}\frac{z_{xx}^{2}-z_y^2}{z_{x}}-3z_x\partial^{-1}\bigg(\bigg(\frac{z_y}{z_x}\bigg)_y\bigg) \nn \\
&& ~~~=z_{t}-z_{xxx}+\frac{3}{2}\frac{z_{xx}^{2}-z_y^2}{z_{x}}-3z_x\bigg(\frac{w_y}{w}-\frac{v_y}{v}\bigg)  \nn \\
&& ~~~=z_{t}-z_{xxx}+\frac{3}{2}\frac{z_{xx}^{2}-z_y^2}{z_{x}}-3(vw_y-wv_y)=0, \label{SKP}
\end{eqnarray}
where $v_y$ and $w_y$ are defined by \eqref{vy} and \eqref{wy}.
It is worth pointing out that \eqref{SKP} is the Schwarzian KP equation \cite{SKP}, which has solution
\begin{equation}
z=\bs^{\st}\bK^{-1}\bC(\bI+\bM\bC)^{-1}\bL^{-1}\br+x,
\end{equation}
where matrices $\bM,~\bK,~\bL$ and vectors $\bs$, $\br$ satisfy
DES \eqref{SE-2} and \eqref{evo-rs}.

Due to the definitions of variables $\mu$ and $\nu$,
relations \eqref{MT-MS}, \eqref{zxx} and \eqref{zy} provide Miura transformations between modified KP equation \eqref{mKP} and
Schwarzian KP equation \eqref{SKP}, respectively, modified KP equation \eqref{an-mKP} and Schwarzian KP equation \eqref{SKP}, given by
\begin{subequations}
\begin{align}
& \mu=\frac{z_{xx}-z_y}{2z_x}, \\
& \nu=\frac{z_{xx}+z_y}{2z_x}.
\end{align}
\label{MT-mKP-SKP}
\end{subequations}

\subsection{The $\tau$-function} \label{tau}

To discuss the bilinear structure of KP system, we introduce the $\tau$-function
\begin{equation}
\tau=|\bI+\bM\bC|,
\label{tau-M}
\end{equation}
for which the following result hold.
\begin{Proposition}\label{prop-5}
For the scalar function $\S{i,j}$ defined in \eqref{Sij} where $\bK, \bL, \bM, \br, \bs$ are formulated by the
Sylvester equation \eqref{SE-2} and $\br, \bs$ obey the evolution \eqref{evo-rs}, we have
\begin{equation}
\S{i,j}=\frac{g}{\tau},
\label{Sij-tau}
\end{equation}
with some function $g=-\left|\begin{array}{cc}
                 0 & \bs^{\st}\,\bK^j\bC\\
                 \bL^i\br & \bI+\bM\bC
                \end{array} \right|$. Specially, for $\S{0,0}$ we have
\begin{equation}
\S{0,0}=\frac{\tau_x}{\tau}.
\label{S00-tau}
\end{equation}
\end{Proposition}
The proof is similar to the one given in \cite{XZZ-2014}.
The rational expressions \eqref{Sij-tau} and \eqref{S00-tau} are always used
in the bilinearization of integrable equations. 

\section{Exact solutions to DES \eqref{SE-2} and \eqref{evo-rs}} \label{sec-3}

Because of the invariance of $\S{i,j}$ under any similar transformation of $\bK$ and $\bL$ (see Sec.\ref{subsubsec-2.2.3}),
here we just need to consider the following simplified/canonical equation set (see \cite{FZ-2013})
\begin{subequations}
\begin{align}
& \Ga\bM-\bM\Lb=\br\bs^{\st}, \label{SE-JCF} \\
& \br_x=\Ga \br,~~\bs_x=-\Lb^{\st} \bs, \label{rs-x-JCF} \\
& \br_y=-\Ga^2 \br,~~\bs_y=(\Lb^{\st})^2 \bs, \label{rs-y-JCF} \\
& \br_t=4\Ga^3 \br,~~\bs_t=-4(\Lb^{\st})^3\bs, \label{rs-t-JCF}
\end{align}
\label{SErs-JCF}
\end{subequations}
where $\Ga$ and $\Lb$ are, respectively, the Jordan canonical forms of $\bL$ and $\bK$.
Corresponding to solvability condition $\mathcal{E}(\bK)\bigcap \mathcal{E}(\bL)=\varnothing$, hereafter we suppose
$\mathcal{E}(\Ga)\bigcap \mathcal{E}(\Lb)=\varnothing$.

\subsection{Some notations} \label{subsec-3.1}

For convenience, some notations will be introduced
where usually the subscripts $_D$ and $_J$ correspond to
the cases of $\Ga$ and $\Lb$ being diagonal and being of Jordan block, respectively, which are listed as follows:

\begin{itemize}
\item{Diagonal matrices:
\begin{subequations}
\begin{align}
& \Ga^{\tyb{N}}_{\ty{D}}(\{l_i\}^{N}_{1})=\mathrm{Diag}(l_1, l_2, \ldots, l_N), \\
& \Lb^{\tyb{N'}}_{\ty{D}}(\{k_j\}^{N'}_{1})=\mathrm{Diag}(k_1, k_2, \cdots, k_{N'}),
\end{align}
\end{subequations}
}
\item{Jordan block matrices:
\begin{subequations}
\begin{align}
& \Ga^{\tyb{N}}_{\ty{J}}(a)
=\left(\begin{array}{cccccc}
a & 0    & 0   & \cdots & 0   & 0 \\
1   & a  & 0   & \cdots & 0   & 0 \\
0   & 1  & a   & \cdots & 0   & 0 \\
\vdots &\vdots &\vdots &\vdots &\vdots &\vdots \\
0   & 0    & 0   & \cdots & 1   & a
\end{array}\right)_{N\times N}, \\
& \Lb^{\tyb{N'}}_{\ty{J}}(b)
=\left(\begin{array}{cccccc}
b & 0    & 0   & \cdots & 0   & 0 \\
1   & b  & 0   & \cdots & 0   & 0 \\
0   & 1  & b   & \cdots & 0   & 0 \\
\vdots &\vdots &\vdots &\vdots &\vdots &\vdots \\
0   & 0    & 0   & \cdots & 1   & b
\end{array}\right)_{N'\times N'},
\end{align}
\end{subequations}
}
\item{Lower triangular Toeplitz matrix: \cite{Zhang-KdV-2006}
\begin{equation}
\bT^{\tyb{N}}(\{a_i\}^{N}_{1})
=\left(\begin{array}{cccccc}
a_1 & 0    & 0   & \cdots & 0   & 0 \\
a_2 & a_1  & 0   & \cdots & 0   & 0 \\
a_3 & a_2  & a_1 & \cdots & 0   & 0 \\
\vdots &\vdots &\cdots &\vdots &\vdots &\vdots \\
a_{N} & a_{N-1} & a_{N-2}  & \cdots &  a_2   & a_1
\end{array}\right)_{N\times N},
\label{T}
\end{equation}
}
\item{Skew triangular Toeplitz matrix:
\begin{equation}
\bH^{\tyb{N'}}(\{b_j\}^{N'}_{1})
=\left(\begin{array}{ccccc}
b_1 & \cdots  & b_{N'-2}  & b_{N'-1} & b_{N'}\\
b_2 & \cdots & b_{N'-1}  & b_{N'} & 0\\
b_3 &\cdots & b_{N'} & 0 & 0\\
\vdots &\vdots & \vdots & \vdots & \vdots\\
b_{N'} & \cdots & 0 & 0 & 0
\end{array}
\right)_{N'\times N'}.
\label{H}
\end{equation}
}
\end{itemize}
Meanwhile, the following expressions need to be considered:
\begin{subequations}\label{notations}
\begin{align}
& \mathrm{exponential~function}: \rho_i=e^{\xi_i},~~\xi_i=l_{i}x-l_i^2y+4l_i^3t+\xi^{(0)}_i,~\mathrm{with~ constants~}\xi^{(0)}_i,\\
& \mathrm{exponential~function}: \sg_i=e^{\eta_i},~~\eta_i=-k_{i}x+k_i^2y-4k_i^3t+\eta^{(0)}_i,~\mathrm{with~ constants~}\eta^{(0)}_i, \\
& N, N'\mathrm{~th\hbox{-}order~vectors:}~~ \br=(r_1, r_2, \cdots, r_N)^{\st},~~  \bs=(s_1, s_2, \cdots, s_{N'})^{\st},\\
& N\times N' ~\mathrm{matrix:}~~\bG^{\tyb{N;N'}}_{\ty{DD}}(\{l_i\}^{N}_{1};\{k_j\}^{N'}_{1})
=(g_{i,j})_{N\times N'},~~~g_{i,j}=\frac{1}{l_i-k_j},\\
& N_1\times N'_2 ~\mathrm{matrix:}~~\bG^{\tyb{N$_1$;N$_2$'}}_{\ty{DJ}}(\{l_i\}^{N_1}_{1};b)
=(g_{i,j})_{N_1\times N'_2},~~~g_{i,j}=\Bigl(\frac{1}{l_i-b}\Bigr)^j,\\
& N_2\times N'_1 ~\mathrm{matrix:}~~\bG^{\tyb{N$_2$;N$_1$'}}_{\ty{JD}}(a;\{k_j\}^{N'_1}_{1})
=(g_{i,j})_{N_2\times N'_1},~~~g_{i,j}=-\Bigl(\frac{-1}{a-k_j}\Bigr)^i,\\
& N_1\times N'_2 ~\mathrm{matrix:}~~\bG^{\tyb{N$_1$;N$_2$'}}_{\ty{JJ}}(a;b)
=(g_{i,j})_{N_1\times N'_2},~~~g_{i,j}=\mathrm{C}^{i-1}_{i+j-2}\frac{(-1)^{i+1}}{(a-b)^{i+j-1}},
\end{align}
\end{subequations}
where \[\mathrm{C}^{i}_{j}=\frac{j!}{i!(j-i)!},~~(j\geq i).\]

\subsection{Exact solutions} \label{subsec-3.2}

The solutions to the Sylvester equation \eqref{SE-2} have been discussed in recent paper
\cite{FZ-2013}, where matrix $\bM$ was factorized into $\bM=\bF \bG \bH$ for $N\times N$ matrix $\bF$, $N\times N'$ matrix $\bG$ and
$N'\times N'$ matrix $\bH$.
Without showing the details, here we just present some main results of solutions to equation set \eqref{SErs-JCF}.

\noindent
Case 1.~ When
\begin{equation}
\Ga=\Ga^{\tyb{N}}_{\ty{D}}(\{l_i\}^{N}_{1}),~~\Lb=\Lb^{\tyb{N'}}_{\ty{D}}(\{k_j\}^{N'}_{1}),
\label{Ga,Lb-DD}
\end{equation}
we have
\begin{subequations}
\begin{align}
&\br=\br_{\hbox{\tiny{\it D}}}^{\hbox{\tiny{[{\it N}]}}}(\{l_i\}_{1}^{N})=(r_1, r_2, \cdots, r_N)^{\st}, ~~\mathrm{with}~ r_i=\rho_i,\\
&\bs=\bs_{\hbox{\tiny{\it D}}}^{\hbox{\tiny{[{\it N'}]}}}(\{k_j\}_{1}^{N'})=(s_1, s_2, \cdots, s_{N'})^{\st}, ~~\mathrm{with}~ s_j=\sigma_j,
\end{align}
\end{subequations}
and
\begin{subequations}
\begin{equation}
\bM  =\bF \bG \bH =\Bigl(\frac{r_i s_j}{l_i-k_j}\Bigr)_{N\times N'}\, ,
\end{equation}
where
\begin{equation}
\bF=\Ga^{\tyb{N}}_{\ty{D}}(\{r_i\}^{N}_{1}),~~
\bG=\bG^{\tyb{N;N'}}_{\ty{DD}}(\{l_i\}^{N}_{1};\{k_j\}^{N'}_{1}),~~
\bH=\Lb^{\tyb{N'}}_{\ty{D}}(\{s_j\}^{N'}_{1}).
\end{equation}
\end{subequations}

\vskip 5pt

\noindent
Case 2.~ When
\begin{equation}
\Ga=\Ga^{\tyb{N}}_{\ty{D}}(\{l_i\}^{N}_{1}),~~\Lb=\Lb^{\tyb{N'}}_{\ty{J}}(k_1),
\label{Ga,Lb-DJ}
\end{equation}
we have
\begin{subequations}
\begin{align}
&\br=\br_{\hbox{\tiny{\it D}}}^{\hbox{\tiny{[{\it N}]}}}(\{l_i\}_{1}^{N})=(r_1, r_2, \cdots, r_N)^{\st}, ~~\mathrm{with}~ r_i=\rho_i,\\
&\bs=\bs_{\hbox{\tiny{\it J}}}^{\hbox{\tiny{[{\it N'}]}}}(k_1)=(s_1, s_2, \cdots, s_{N'})^{\st},
~~\mathrm{with}~ s_j=\frac{\partial^{N'-j}_{k_1}\sigma_1}{(N'-j)!},
\end{align}
\end{subequations}
and
\begin{subequations}
\begin{equation}
\bM=\bF  \bG  \bH,
\end{equation}
where
\begin{equation}
\bF=\Ga^{\tyb{N}}_{\ty{D}}(\{r_i\}^{N}_{1}),~~\bG=\bG^{\tyb{N;N'}}_{\ty{DJ}}(\{l_i\}_1^N;k_1),~~ \bH=\bH^{\tyb{N'}}(\{s_j\}^{N'}_{1}).
\end{equation}
\end{subequations}
Likewise, we can also obtain the solution for equation set (3.6) when
$\Ga=\Ga^{\tyb{N}}_{\ty{J}}(l_{1}),~~\Lb=\Lb^{\tyb{N'}}_{\ty{D}}(\{k_j\}^{N'}_{1})$.

\vskip 5pt
\noindent
Case 3.~When
\begin{equation}
\Ga=\Ga^{\tyb{N}}_{\ty{J}}(l_1),~~\Lb=\Lb^{\tyb{N'}}_{\ty{J}}(k_1),
\label{Ga,Lb-JJ}
\end{equation}
we have
\begin{subequations}
\begin{align}
&\br=\br_{\hbox{\tiny{\it J}}}^{\hbox{\tiny{[{\it N}]}}}(l_1)=(r_1, r_2, \cdots, r_N)^{\st},
~~\mathrm{with}~ r_i=\frac{\partial^{i-1}_{l_1}\rho_1}{(i-1)!},\\
&\bs=\bs_{\hbox{\tiny{\it J}}}^{\hbox{\tiny{[{\it N'}]}}}(k_1)=(s_1, s_2, \cdots, s_{N'})^{\st},
~~\mathrm{with}~ s_j=\frac{\partial^{N'-j}_{k_1}\sigma_1}{(N'-j)!},
\end{align}
\end{subequations}
and
\begin{subequations}
\begin{equation}
\bM=\bF  \bG  \bH,
\end{equation}
where
\begin{equation}
\bF=\bT^{\tyb{N}}(\{r_i\}^{N}_{1}),~~\bG=\bG^{\tyb{N;N'}}_{\ty{JJ}}(l_1;k_1),~~ \bH=\bH^{\tyb{N'}}(\{s_j\}^{N'}_{1}).
\end{equation}
\end{subequations}

\vskip 5pt
\noindent
Case 4.~When $\Ga$ and $\Lb$ are taken as
\begin{subequations}
\label{Ga,Lb-gen}
\begin{align}
& \Ga=\mathrm{Diag}\bigl(\Ga^{\tyb{N$_1$}}_{\ty{D}}(\{l_i\}^{N_1}_{1}),
\Ga^{\tyb{N$_2$}}_{\ty{J}}(l_{N_1+1}),\Ga^{\tyb{N$_3$}}_{\ty{J}}(l_{N_1+2}),\cdots,
\Ga^{\tyb{N$_s$}}_{\ty{J}}(l_{N_1+(s-1)})\bigr), \\
& \Lb=\mathrm{Diag}\bigl(\Lb^{\tyb{N$_1$'}}_{\ty{D}}(\{k_j\}^{N'_1}_{1}),
\Lb^{\tyb{N$_2$'}}_{\ty{J}}(k_{N'_1+1}),\Lb^{\tyb{N$_3$'}}_{\ty{J}}(k_{N'_1+2}),\cdots,
\Lb^{\tyb{N$_s$'}}_{\ty{J}}(k_{N'_1+(s-1)})\bigr),
\end{align}
\end{subequations}
with $\sum_{i=1}^sN_i=N$ and $\sum_{i=1}^sN'_i=N'$,
then we have solutions
\begin{equation}
\br=\left(
\begin{array}{l}
\br_{\ty{D}}^{\tyb{N$_1$}}(\{l_i\}_{1}^{N_1})\\
\br_{\ty{J}}^{\tyb{N$_2$}}(l_{N_1+1})\\
\br_{\ty{J}}^{\tyb{N$_3$}}(l_{N_1+2})\\
\vdots\\
\br_{\ty{J}}^{\tyb{N$_s$}}(l_{N_1+(s-1)})
\end{array}
\right),~~~\bs=\left(
\begin{array}{l}
\bs_{\ty{D}}^{\tyb{N$_1$'}}(\{k_j\}_{1}^{N'_1})\\
\bs_{\ty{J}}^{\tyb{N$_2$'}}(k_{N'_1+1})\\
\bs_{\ty{J}}^{\tyb{N$_3$'}}(k_{N'_1+2})\\
\vdots\\
\bs_{\ty{J}}^{\tyb{N$_s$'}}(k_{N'_1+(s-1)})
\end{array}
\right),
\end{equation}
and $\bM=\bF\bG \bH$, where
\begin{align}
&\bF=\mathrm{Diag}\bigl(
\Ga^{\tyb{N$_1$}}_{\ty{D}}(\{r_i\}^{N_1}_{1}),
\bT^{\tyb{N$_2$}}(l_{N_1+1}),\bT^{\tyb{N$_3$}}(l_{N_1+2}),\cdots,
\bT^{\tyb{N$_s$}}(l_{N_1+(s-1)})
\bigr),\label{KP-r-M-g-F}\\
&\bH=\mathrm{Diag}\bigl(
\Lb^{\tyb{N$_1$'}}_{\ty{D}}(\{s_j\}^{N'_1}_{1}),
\bH^{\tyb{N$_2$'}}(k_{N'_1+1}),
\bH^{\tyb{N$_3$'}}(k_{N'_1+2}),
\cdots,
\bH^{\tyb{N$_s$'}}(k_{N'_1+(s-1)})\bigr),\label{KP-r-M-g-H}
\end{align}
and $\bG$ possesses block structure
\begin{equation}
\bG=(\bG_{i,j})_{s\times s},
\label{KP-r-M-g-G1}
\end{equation}
with
\begin{subequations}\label{KP-r-M-g-G2}
\begin{align}
& \bG_{1,1}=\bG^{\tyb{N$_1$;N$_1$'}}_{\ty{DD}}(\{l_i\}^{N_1}_{1};\{k_j\}^{N'_1}_{1}), \label{KP-bs-sol-1}\\
& \bG_{1,j}=\bG^{\tyb{N$_1$;N$_j$'}}_{\ty{DJ}}(\{l_i\}^{N_1}_{1};k_{N'_1+j-1}),~~~(1<j\leq s), \label{KP-bs-sol-2}\\
& \bG_{i,1}=\bG^{\tyb{N$_i$;N$_1$'}}_{\ty{JD}}(l_{N_1+i-1};\{k_j\}^{N'_1}_{1}),~~~(1<i\leq s), \label{KP-bs-sol-3}\\
& \bG_{i,j}=\bG^{\tyb{N$_i$;N$_j$'}}_{\ty{JJ}}(l_{N_1+i-1};k_{N'_1+j-1}),~~~(1<i,j\leq s). \label{KP-bs-sol-4}
\end{align}
\end{subequations}

In summary, in this section we have derived all explicit solutions for the DES \eqref{SErs-JCF}.
By master function $\S{i,j}$, these solutions lead to various type of solutions for KP system.
The multi-soliton solutions and the general mixed solutions(soliton-Jordan block mixed solutions)
are obtained in Case 1 and Case 4, respectively.
The Jordan block solution in Case 3 corresponds to the multiple-pole solutions, which can be obtained from
multi-soliton solutions of Case 1 through limit procedure by taking $\{l_j\}_{j=2}^N\rightarrow l_1$
and $\{k_j\}_{j=2}^{N'}\rightarrow k_1$, successively. In Case 2, since $\Ga$ is a diagonal
and $\Lb$ is a Jordan-block, the solution given in this case has both properties of soliton and multiple-pole solution.
Since we assume $\Ga$ and $\Lb$ satisfy invertible conditions,
eigenvalues of $\Ga$ and $\Lb$ can not be zero and
the obtained solutions here do not include rational solutions.

\section{Reduction} \label{sec-4}

As we all know, KdV equation and BSQ equation can be derived from
KP equation by imposing dimensional reductions. In this section, the reduction of the obtained KP system
given in Sec. \ref{subsec-2.3} will be carried out by taking constraints on matrices $\bK$ and
$\bL$ in DES \eqref{SE-2} and \eqref{evo-rs}. As a consequence, KdV system, BSQ system and extended BSQ system will be
obtained. For this purpose, we take $N'=N$ in DES \eqref{SE-2} and \eqref{evo-rs} while $\bC=\bI$ in
scalar function $\S{i,j}$.

\subsection{Reduction to KdV system} \label{sec-4.1}

The generalized Cauchy matrix approach for KdV system has been discussed in Ref. \cite{XZZ-2014}. To derive the KdV
system from the KP system by reduction,
we suppose $\bK=-\bL$. Then DES \eqref{SE-2} and \eqref{evo-rs}
lead to equation set
\begin{subequations}
\begin{align}
& \bL \bM+\bM\bL=\br\, \bs^{\st}, \label{Syl-KdV}\\
& \br_x=\bL \br,~~\bs_x=\bL^{\st} \bs, \label{evo-rs-x-KdV} \\
& \br_y=-\bL^2 \br,~~\bs_y=(\bL^{\st})^2 \bs, \label{evo-rs-y-KdV} \\
& \br_t=4\bL^3 \br,~~\bs_t=4(\bL^{\st})^3\bs, \label{evo-rs-t-KdV}
\end{align}
\label{evo-rs-KdV}
\end{subequations}
and $\S{i,j}$ in \eqref{Sij} is of form
\begin{align}
\S{i,j}=(-1)^j\bs^{\st}\,\bL^j(\bI+\bM)^{-1}\bL^i\br. \label{Sij-KdV}
\end{align}
Comparing the equality \eqref{Sij-2k=1+} with \eqref{evo-Sijy},
we have $S^{(i,j)}_y=0$. Hence KdV system can be obtained immediately from
KP system \eqref{KP}, \eqref{mKP}, \eqref{an-mKP} and
\eqref{SKP}. We list these equations as follows:

\vspace{.3cm}

\noindent{\it {KdV equation}:}
\begin{eqnarray}
\va_{t}-\va_{xxx}-6\va\va_{x}=0, \label{KdV}
\end{eqnarray}
of which the solution is given by
\begin{equation}
\va=2\S{0,0}_x=2(\bs^{\st}(\bI+\bM)^{-1}\br)_x. \label{KdV-so}
\end{equation}

\vspace{.3cm}

\noindent{\it {modified KdV equation}:} It follows from the anti-symmetric property $S^{(-1,0)}=-S^{(0,-1)}$ that
$\mu=\nu=\partial_x\ln(1-S^{(-1,0)})$. Thus both modified KP equations \eqref{mKP} and \eqref{an-mKP} reduce to the same modified KdV equation
\begin{eqnarray}
\mu_{t}-\mu_{xxx}+6\mu^2\mu_x=0, \label{mKdV}
\end{eqnarray}
of which the solution is given by
\begin{equation}
\mu=\partial_x\ln(1-S^{(-1,0)})=\partial_x \ln (1-\bs^{\st}(\bI+\bM)^{-1}\bL^{-1}\br). \label{mKdV-so}
\end{equation}
A natural fact is that \eqref{MT-KP-mKP} and \eqref{MT-KP-an-mKP} reduce to the Miura transformation
between modified KdV equation \eqref{mKdV} and KdV equation \eqref{KdV}, i.e.
\begin{eqnarray}
-\va=\mu_x+\mu^2.
\end{eqnarray}

\vspace{.3cm}

\noindent{\it {Schwarzian KdV equation}:}
\begin{eqnarray}
z_{t}-z_{xxx}+\frac{3}{2}\frac{z_{xx}^{2}}{z_{x}}=0, \label{SKdV}
\end{eqnarray}
of which the solution is given by
\begin{equation}
z=\S{-1,-1}+x=-\bs^{\st}\bL^{-1}(\bI+\bM)^{-1}\bL^{-1}\br+x.  \label{SKdV-so}
\end{equation}
In this case, Miura transformation
between modified KdV equation \eqref{mKdV} and Schwarzian KdV equation \eqref{SKdV} is described as
\[\mu=\frac{z_{xx}}{2z_x}.\]
In solutions \eqref{KdV-so}, \eqref{mKdV-so} and \eqref{SKdV-so}, $\bL,~\bM,~\br$ and $\bs$
are determined by equation set \eqref{evo-rs-KdV}.

\subsection{Reduction to BSQ system} \label{sec-4.2}
\vspace{.3cm}

Now the reduction to BSQ system will be carried out due to different relations between $\bL$ and $\bK$.
We consider the following two cases:

\noindent{\bf \underline{Case 1}:} $\bK=\oa\bL$, where $\oa$ is defined in Proposition \ref{prop-BSQ-1}.
In this case, DES \eqref{SE-2} and \eqref{evo-rs} become
\begin{subequations}
\begin{align}
& \bL \bM-\oa\bM\bL=\br\, \bs^{\st}, \label{Syl-BSQ}\\
& \br_x=\bL \br,~~\bs_x=-\oa\bL^{\st} \bs, \label{evo-rs-x-BSQ} \\
& \br_y=-\bL^2 \br,~~\bs_y=\oa^2(\bL^{\st})^2 \bs, \label{evo-rs-y-BSQ} \\
& \br_t=4\bL^3 \br,~~\bs_t=-4(\bL^{\st})^3\bs, \label{evo-rs-t-BSQ}
\end{align}
\label{evo-rs-BSQ}
\end{subequations}
and $\S{i,j}$ in \eqref{Sij} reads
\begin{align}
\S{i,j}=\oa^j\bs^{\st}\,\bL^j(\bI+\bM)^{-1}\bL^i\br. \label{Sij-KdV}
\end{align}
Noting that the equality \eqref{Sij-3k=1+} and the expression of
$S^{(i,j)}_t$ given in \eqref{evo-Sijt}, we find $S^{(i,j)}_t=0$.
The KP system \eqref{KP}, \eqref{mKP}, \eqref{an-mKP} and
\eqref{SKP} thereby reduce to the BSQ system. We list these equations as follows, where
$y$ can be viewed as time variable.

\vspace{.3cm}

\noindent{\it {BSQ equation}:}
\begin{eqnarray}
\va_{xxx}+6\va\va_{x}+3\partial^{-1}\va_{yy}=0, \label{BSQ}
\end{eqnarray}
of which the solution is given by
\begin{equation}
\va=2\S{0,0}_x=2(\bs^{\st}(\bI+\bM)^{-1}\br)_x. \label{BSQ-so}
\end{equation}

\vspace{.3cm}

\noindent{\it {modified BSQ equation}:} modified KP equation \eqref{mKP}
reduces to the modified BSQ equation
\begin{eqnarray}
-\mu_{xxx}+6\mu^2\mu_x+6\mu_x\partial^{-1}\mu_y-3\partial^{-1}\mu_{yy}=0, \label{mBSQ}
\end{eqnarray}
of which the solution is given by
\begin{equation}
\mu=\partial_x \ln(1+\S{0,-1})=\partial_x \ln (1+\frac{1}{\oa}\bs^{\st}\bL^{-1}(\bI+\bM)^{-1}\br). \label{mBSQ-so}
\end{equation}
In this case, \eqref{MT-KP-mKP} becomes the Miura transformation
between modified BSQ equation \eqref{mBSQ} and BSQ equation \eqref{BSQ}.

Similarly, \eqref{an-mKP} reduces to another modified BSQ equation
\begin{eqnarray}
-\nu_{xxx}+6\nu^2\nu_x-6\nu_x\partial^{-1}\nu_y-3\partial^{-1}\nu_{yy}=0, \label{an-mBSQ}
\end{eqnarray}
of which the solution is given by
\begin{equation}
\nu=\partial_x \ln(1-\S{-1,0})=\partial_x \ln (1-\bs^{\st}(\bI+\bM)^{-1}\bL^{-1}\br). \label{an-mBSQ-so}
\end{equation}
Naturally, \eqref{MT-KP-an-mKP} turns into the Miura transformation
between modified BSQ equation \eqref{an-mBSQ} and BSQ equation \eqref{BSQ}.

\vspace{.3cm}

\noindent{\it {Schwarzian BSQ equation}:}
\begin{eqnarray}
-z_{xxx}+\frac{3}{2}\frac{z_{xx}^{2}-z_y^2}{z_{x}}-3z_x\partial^{-1}\bigg(\bigg(\frac{z_y}{z_x}\bigg)_y\bigg)=0, \label{SBSQ}
\end{eqnarray}
of which the solution is given by
\begin{equation}
z=\S{-1,-1}+x=\frac{1}{\oa}\bs^{\st}\bL^{-1}(\bI+\bM)^{-1}\bL^{-1}\br+x. \label{SBSQ-so}
\end{equation}
The Miura transformations
between modified BSQ equation \eqref{mBSQ} and Schwarzian BSQ equation \eqref{SBSQ}, respectively,
modified BSQ equation \eqref{an-mBSQ} and Schwarzian BSQ equation \eqref{SBSQ}, are still \eqref{MT-mKP-SKP}.
In \eqref{BSQ-so}, \eqref{mBSQ-so}, \eqref{an-mBSQ-so} and \eqref{SBSQ-so}, $\bL,~\bM,~\br$ and $\bs$
are determined by equation set \eqref{evo-rs-BSQ}.

\noindent{\bf \underline{Case 2}:} Supposing
\begin{align}
\bK=\left(
\begin{array}{cc}
\oa\bK_1 & \bm 0  \\
\bm 0 & \oa^2\bK_2
\end{array}\right),~~\bL=\left(
\begin{array}{cc}
\bK_1 & \bm 0  \\
\bm 0 & \bK_2
\end{array}\right), \label{R2-BSQ}
\end{align}
where $\bK_i \in \mathbb{C}_{N_i\times N_i}$ with $N_1+N_2=N$,
then for the object \eqref{Sij} defined by
DES \eqref{SE-2} and \eqref{evo-rs}, we have the Proposition \ref{prop-BSQ-2},
which implies $S^{(i,j)}_t=0$. Consequently, KP system \eqref{KP}, \eqref{mKP}, \eqref{an-mKP} and
\eqref{SKP} still, respectively, lead to BSQ equation \eqref{BSQ}, modified BSQ equations \eqref{mBSQ},
\eqref{an-mBSQ} and Schwarzian BSQ equation \eqref{SBSQ}.

Although the reduced BSQ system of Case 1 is the same as the one of Case 2, the exact solutions
for these two cases are totally different. Similar to discrete case \cite{ZZS-2012,FZZ-2012}, in Case 1
$\S{i,j}$ only contains one kind of plain wave factor while in Case 2 $\S{i,j}$ has two kinds of
plain wave factors. To make a comparison,
1-soliton solution corresponding to these two
cases are presented in Appendix \ref{A:SS}.

\subsection{Reduction to extended BSQ system} \label{sec-4.3}

The discrete extended BSQ system was firstly proposed by Hietarinta in \cite{H-2011}
and studied systematically by Zhang {\it et al.} applying direct linearization method \cite{ZZN-2012}.
Here we will construct the continue extended BSQ system, which can be
viewed as the continuous correspondence of the discrete extended BSQ system given in \cite{H-2011}.

Assume
\begin{align}
\bK=\left(
\begin{array}{cc}
\boa_1(\bK_1) & \bm 0  \\
\bm 0 & \boa_2(\bK_2)
\end{array}\right),~~\bL=\left(
\begin{array}{cc}
\bK_1 & \bm 0  \\
\bm 0 & \bK_2
\end{array}\right), \label{ex-KL}
\end{align}
where $\boa_1(\bK_1)$ and $\boa_2(\bK_2)$ are defined by \eqref{eq:g}.
With the help of the equality \eqref{Sij-3alk=1+} and the expressions of
$S_x^{(i,j)}$ and $S_y^{(i,j)}$, we have
\begin{align}
& S_t^{(i,j)}=4\al_2(-S^{(i+2,j)}+S^{(i,j+2)}+S^{(i,1)}S^{(0,j)}+S^{(i,0)}S^{(1,j)}) \nn \\
&~~~~~~~~~+4\al_1(S^{(i,j+1)}-S^{(i+1,j)}+S^{(i,0)}S^{(0,j)}) \nn \\
&~~~~~~=4(\al_2S_y^{(i,j)}-\al_1S_x^{(i,j)}). \label{eB-eq}
\end{align}
By virtue of \eqref{eB-eq}, KP system \eqref{KP}, \eqref{mKP}, \eqref{an-mKP} and
\eqref{SKP} give rise to the extended BSQ-type equations. The results are listed as follows:

\vspace{.3cm}

\noindent{\it {extended BSQ equation}:}
\begin{eqnarray}
\va_{xxx}+6\va\va_{x}+3\partial^{-1}\va_{yy}-4(\al_2\va_y-\al_1\va_x)=0, \label{ex-BSQ}
\end{eqnarray}
of which the solution is given by
\begin{equation}
w=2\S{0,0}_x=2(\bs^{\st}(\bI+\bM)^{-1}\br)_x.  \label{ex-BSQ-so}
\end{equation}

\vspace{.3cm}

\noindent{\it {extended modified BSQ equation}:} The extended modified BSQ equation
reduced from modified KP equation \eqref{mKP} reads
\begin{eqnarray}
-\mu_{xxx}+6\mu^2\mu_x+6\mu_x\partial^{-1}\mu_y-3\partial^{-1}\mu_{yy}
+4(\al_2\partial^{-1}\mu_y-\al_1\mu)=0, \label{ex-mBSQ}
\end{eqnarray}
of which the solution is given by
\begin{equation}
\mu=\partial_x \ln (1+\S{0,-1})=\partial_x \ln (1+\bs^{\st}\bK^{-1}(\bI+\bM)^{-1}\br). \label{ex-mBSQ-so}
\end{equation}
In this case, \eqref{MT-KP-mKP} becomes the Miura transformation
between extended modified BSQ equation \eqref{ex-mBSQ} and extended BSQ equation \eqref{ex-BSQ}.

Similarly, \eqref{an-mKP} reduces to another extended modified BSQ equation
\begin{eqnarray}
-\nu_{xxx}+6\nu^2\nu_x-6\nu_x\partial^{-1}\nu_y-3\partial^{-1}\nu_{yy}+4(\al_2\partial^{-1}\nu_y-\al_1\nu)=0, \label{an-ex-mBSQ}
\end{eqnarray}
of which the solution is given by
\begin{equation}
\nu=\partial_x \ln (1-\S{-1,0})=\partial_x \ln (1-\bs^{\st}(\bI+\bM)^{-1}\bL^{-1}\br).  \label{an-ex-mBSQ-so}
\end{equation}
\eqref{MT-KP-an-mKP} becomes the Miura transformation
between extended modified BSQ equation \eqref{an-ex-mBSQ} and extended BSQ equation \eqref{BSQ}.

\vspace{.3cm}

\noindent{\it {extended Schwarzian BSQ equation}:}
\begin{eqnarray}
-z_{xxx}+\frac{3}{2}\frac{z_{xx}^{2}-z_y^2}{z_{x}}-3z_x\partial^{-1}\bigg(\bigg(\frac{z_y}{z_x}\bigg)_y\bigg)
+4(\al_2z_y-\al_1z_x+\al_1)=0,
\label{ex-SBSQ}
\end{eqnarray}
of which the solution is given by
\begin{equation}
z=\S{-1,-1}+x=\bs^{\st}\bK^{-1}(\bI+\bM)^{-1}\bL^{-1}\br+x.  \label{ex-SBSQ-so}
\end{equation}
Similarly, \eqref{MT-mKP-SKP} turns into the Miura transformations
between extended modified BSQ equation \eqref{ex-mBSQ} and extended Schwarzian BSQ equation \eqref{ex-SBSQ}, respectively,
extended modified BSQ equation \eqref{an-ex-mBSQ} and extended Schwarzian BSQ equation \eqref{ex-SBSQ}.

In \eqref{ex-BSQ-so}, \eqref{ex-mBSQ-so}, \eqref{an-ex-mBSQ-so} and \eqref{ex-SBSQ-so}, $\bM,~\br$ and $\bs$
are determined by DES \eqref{SE-2} and \eqref{evo-rs}, where $\bL$ and $\bK$ are given by \eqref{ex-KL}.

\section{Conclusions} \label{Con}

Generalized Cauchy matrix approach can be viewed as a direct generalization of the
Cauchy matrix approach ($\bK$ and $\bL$ in \eqref{SE-2} and \eqref{evo-rs} are known diagonal matrices
which lead to soliton solutions). This method has close connection with the
direct linearization method \cite{AS-book,FA-KP,FA-KP-1}. Both the direct linearization method and the generalized Cauchy matrix approach
can be used to yield a system of equations as well as broad kinds of solutions besides the inverse scattering type solutions.
In this paper, the generalized Cauchy matrix approach
is applied to study the continue KP system, where KP equation, modified KP equation and Schwarzian KP equation are constructed,
whose solutions are expressed by scalar function $\S{i,j}$. By setting different forms of $\Ga$ and $\Lb$ in canonical
equation set \eqref{SErs-JCF}, various solutions are obtained, including solitons, Jordan-block solutions and mixed solutions.
The procedure shown in present paper can be viewed
as a continuous version of the generalized Cauchy matrix approach in the discrete case \cite{N-2004-math}. By imposing some
constrains on $\bK$ and $\bL$, many recurrence relations of $\S{i,j}$ have been derived (See Propositions \ref{prop-KdV}-\ref{prop-exBSQ}).
In terms of these identities, KdV system, BSQ system and extended BSQ system are constructed. The
extended BSQ system \eqref{ex-BSQ}, \eqref{ex-mBSQ}, \eqref{an-ex-mBSQ} and \eqref{ex-SBSQ}
can be viewed as continuous version of the discrete extended BSQ system given in
\cite{H-2011}.

When vectors $\br$ and $\bs$ in DES \eqref{SE-2} and \eqref{evo-rs} are replaced with matrices,
then matrix KP system or noncommutative KP system can be constructed,
which would be considered in future.
By means of Cauchy matrix approach, elliptic soliton solutions to lattice KdV system, ABS lattice and
lattice KP system have been constructed by Nijhoff
and his collaborators in recent papers \cite{E-ABS,E-KP}. It is of great interest to discuss
the elliptic soliton solutions for continue integrable system by utilizing this method,
which will be one part of ongoing researches.

\vskip 20pt
\subsection*{Acknowledgments}
The first author thanks Prof. Da-jun Zhang for the enthusiastic discussion.
This project is supported by the National Natural Science Foundation
Grant (Nos. 11301483, 11401529 and 11371323).

\vskip 20pt

\begin{appendix}

\section{Algebraic relation with extended BSQ system} \label{A:G3}

In \cite{ZZN-2012}, Zhang {\it et al.} introduced an algebraic relation
\begin{eqnarray}
\label{eq:G3-def}
G_{3}(\oa,k):=g(\oa)-g(k)=0,~{\rm where}~g(k)=\sum_{j=1}^{3}\al_jk^{j},~~\al_{3}=1,
\end{eqnarray}
whose roots are denoted by $\oa_j(k)(j=1,2,3)$, i.e.
\begin{subequations}
\begin{eqnarray}
&& \oa_1(k)=\frac{1}{2}\big(-\al_2-k+\sqrt{(\al_2-3k)(\al_2+k)-4\al_1}\big), \\
&& \oa_2(k)=\frac{1}{2}\big(-\al_2-k-\sqrt{(\al_2-3k)(\al_2+k)-4\al_1}\big), \\
&& \oa_3(k)=k.
\end{eqnarray}
\label{eq:oa-123}
\end{subequations}
Obviously these solutions satisfy relations
\begin{subequations}
\begin{eqnarray}
&& \oa_1(k)+ \oa_2(k)+ \oa_3(k)=-\al_2, \\
&& \oa_1(k)\oa_2(k)+\oa_1(k)\oa_3(k)+\oa_2(k)\oa_3(k)=\al_1, \\
&& \oa_1(k)\oa_2(k)\oa_3(k)=k^3+\al_2k^2+\al_1k.
\end{eqnarray}
\label{eq:oa-123-re}
\end{subequations}
By \eqref{eq:oa-123} one can define $N\times N$ matrices $\boa_j(\bK)(j=1,2,3)$ (See Ref.\cite{Z-INI-2013}):
\begin{subequations}
\begin{eqnarray}
&& \boa_1(\bK)=\frac{1}{2}\big(-\al_2\bI-\bK+\sqrt{(\al_2\bI-3\bK)(\al_2\bI+\bK)-4\al_1\bI}\big), \\
&& \boa_2(\bK)=\frac{1}{2}\big(-\al_2\bI-\bK-\sqrt{(\al_2\bI-3\bK)(\al_2\bI+\bK)-4\al_1\bI}\big), \\
&& \boa_3(\bK)=\bK.
\end{eqnarray}
\label{eq:moa-123}
\end{subequations}
It is easy to know that \eqref{eq:moa-123} are roots of the following
matrix equation
\begin{eqnarray}
\label{eq:mG3-def}
G_{3}(\boa,\bK):=g(\boa)-g(\bK)=0,~{\rm where}~g(\bK)=\sum_{j=1}^{3}\al_j\bK^{j},~~\al_{3}=1.
\end{eqnarray}
Now we set
\begin{eqnarray}
\label{eq:KL}
\bL=\mbox{Diag}(\bK_1,\bK_2),~~
\bK=\mbox{Diag}(\boa_1(\bK_1),\boa_2(\bK_2)),
\end{eqnarray}
with $\bK_j \in \mathbb{C}_{N_j \times N_j}$ and $N_1+N_2=N$. Then it is easy to know that
\begin{eqnarray}
\label{eq:equa}
\prod_{h=1}^{3}(\oa_h(a)\bI-\bL)=\prod_{h=1}^{3}(\oa_h(a)\bI-\bK)
\end{eqnarray}
holds for arbitrary constant $a$. Noting that \eqref{eq:equa}, for matrix $\bM$ satisfying the Sylvester equation \eqref{SE-2}
with $\bL$ and $\bK$ defined by \eqref{eq:KL}, we have the following equality
\begin{eqnarray}
&& \prod_{h=1}^{3}(\oa_h(a)\bI-\bK)\bM=\bM\prod_{h=1}^{3}(\oa_h(a)\bI-\bK) \nn \\
&&~~~~~~~~~~~~~~~~~~~~~~~~~~~~~~
-\sum_{k=1}^{3}[\prod_{h=1}^{k-1}(\oa_h(a)\bI-\bL)]\br\bs^{\st}[\prod_{h=k+1}^{3}(\oa_h(a)\bI-\bK)].
\label{eq:M-eq1}
\end{eqnarray}
Left-multiplying $(\bI+\bM)\bu^{(i)}=\bL^i\br$ by term
$\prod_{h=1}^{3}(\oa_h(a)\bI-\bK)$ and making use of \eqref{eq:M-eq1} yield
\begin{eqnarray}
&& (\bI+\bM)\prod_{h=1}^{3}(\oa_h(a)\bI-\bK)\bu^{(i)}=\prod_{h=1}^{3}(\oa_h(a)\bI-\bL)\bL^i\br \nn \\
&&~~~~~~~~~~~~~~~~~~~~~~~~
~~~+\sum_{k=1}^{3}[\prod_{h=1}^{k-1}(\oa_h(a)\bI-\bL)]\br\bs^{\st}[\prod_{h=k+1}^{3}(\oa_h(a)\bI-\bK)]\bu^{(i)}.
\label{ui-dyn}
\end{eqnarray}
Multiplying \eqref{ui-dyn} by $\bs^{\st}\bK^j$ from the left associated with \eqref{evo-rs} and
\eqref{Sij-ui}, we get the relation for scalar function
$S^{(i,j)}=\bs^{\st}\bK^j(\bI+\bM)^{-1}\bL^i\br$, which is given by
\begin{eqnarray}
&& S^{(i+3,j)}-S^{(i,j+3)}-S^{(i,2)}S^{(0,j)}-S^{(i,1)}S^{(1,j)}-S^{(i,0)}S^{(2,j)} \nn \\
&&~~~~= \al_2(S^{(i,j+2)}-S^{(i+2,j)}+S^{(i,0)}S^{(1,j)}+S^{(i,1)}S^{(0,j)}) \nn \\
&& ~~~~~~~+\al_1(S^{(i,j+1)}-S^{(i+1,j)}+S^{(i,0)}S^{(0,j)}),
\label{App-eq}
\end{eqnarray}
where we have utilized relation \eqref{eq:oa-123-re}. Equality \eqref{App-eq} is \eqref{Sij-3alk=1+} in Proposition \ref{prop-exBSQ}.

%

\section{1-soliton solution to BSQ system} \label{A:SS}

For Case 1 in Sec. \ref{sec-4.2}, we take $N=1$ and $\bL=l$. Solving \eqref{evo-rs-BSQ}
yields
\begin{subequations}
\begin{align}
\br=e^{\xi},~~\bs=e^{\eta},~~\bM=\frac{e^{\xi+\eta}}{(1-\oa)l},
\end{align}
where
\begin{align}
& \xi=lx-l^2y+4l^3t+\xi^{(0)}, \\
& \eta=-\oa lx+\oa^2l^2y-4l^3t+\eta^{(0)}
\end{align}
\end{subequations}
with $\xi^{(0)},~\eta^{(0)} \in \mathbb{C}$.
In this case, $S^{(i,j)}$ is of form
\begin{align}
S^{(i,j)}=\oa^jl^{i+j+1}\frac{(1-\oa)e^{\zeta}}{(1-\oa)l+e^{\zeta}},
\label{soli-1}
\end{align}
where $\zeta=(1-\oa)lx-(1-\oa^2)l^2y+\zeta^{(0)}$ with $\zeta^{(0)}=\xi^{(0)}+\eta^{(0)}$.

For Case 2 in Sec. \ref{sec-4.2}, we take $N=2$ and
\[
\bK=\left(
\begin{array}{cc}
\oa l &  0  \\
 0 & \oa^2 l
\end{array}\right),~~\bL=\left(
\begin{array}{cc}
l & 0  \\
0 & l
\end{array}\right).
\]
Solving DES \eqref{SE-2} and \eqref{evo-rs} leads to
\begin{subequations}
\begin{align}
\br=\left(
\begin{array}{c}
e^{\xi_1} \\
e^{\xi_2}
\end{array}\right),~~\bs=\left(
\begin{array}{c}
e^{\eta_1} \\
e^{\eta_2}
\end{array}\right), \label{rs-so}
\end{align}
and
\begin{align}
\bM=\left(
\begin{array}{cc}
\frac{e^{\xi_1+\eta_1}}{l(1-\oa)} & \frac{e^{\xi_1+\eta_2}}{l(1-\oa^2)} \\
\frac{e^{\xi_2+\eta_1}}{l(1-\oa)} & \frac{e^{\xi_2+\eta_2}}{l(1-\oa^2)}
\end{array}\right), \label{M-so}
\end{align}
where
\begin{align}
& \xi_i=lx-l^2y+4l^3t+\xi_i^{(0)}, \\
& \eta_1=-\oa lx+\oa^2l^2y-4l^3t+\eta_1^{(0)}, \\
& \eta_2=-\oa^2 lx+\oa l^2y-4l^3t+\eta_2^{(0)}
\end{align}
\end{subequations}
with $\xi_i^{(0)},~\eta_i^{(0)} \in \mathbb{C},~~(i=1,2)$.
Substituting \eqref{rs-so} and \eqref{M-so} into $S^{(i,j)}=\bs^{\st}\bK^j(\bI+\bM)^{-1}\bL^i\br$,
we arrive at
\begin{align}
S^{(i,j)}=3l^{i+j+1}\frac{\oa^je^{\zeta_1}
+\oa^{2j}e^{\zeta_2}}{3l+(1-\oa^2)le^{\zeta_1}
+(1-\oa)le^{\zeta_2}},
\label{soli-2}
\end{align}
where
\begin{align}
& \zeta_1=(1-\oa)lx-(1-\oa^2)l^2y+\zeta_1^{(0)}, \\
& \zeta_2=(1-\oa^2)lx-(1-\oa)l^2y+\zeta_2^{(0)}
\end{align}
with $\zeta_i^{(0)}=\xi_i^{(0)}+\eta_i^{(0)},~~(i=1,2)$.

It is worthy to note that solutions \eqref{soli-1} and \eqref{soli-2} don't contain variable $t$, which coincide with $S^{(i,j)}_t=0$
given in subsection \ref{sec-4.2}.
Although both \eqref{soli-1} and \eqref{soli-2} give 1-soliton solution for BSQ system via
the corresponding transformations,
only one plain wave factor $e^{\zeta}$ contains in \eqref{soli-1} and two
plain wave factors $e^{\zeta_1}$ and $e^{\zeta_2}$ appear in \eqref{soli-2}.

\end{appendix}

\end{document}